\documentclass[12pt]{article}
\usepackage{graphicx}

\def\a{\alpha}
\def\b{\beta}

\def\f{\phi}
\def\g{\gamma}
\def\h{\eta}

\def\j{\psi}

\def\l{\lambda}
\def\m{\mu}
\def\n{\nu}

\def\p{\pi}

\def\s{\sigma}
\def\t{\tau}

\def\D{\Delta}
\def\F{\Phi}

\def\L{\Lambda}

\def\Q{\Theta}
\def\S{\Sigma}

\def\ve{\varepsilon}

\def\tm{\widetilde{\mu}}

\def\tM{\widetilde{M}}

\def\tM{\widetilde{M}}

\def\beq{\begin{equation}}
\def\eeq{\end{equation}}
\def\bea{\begin{eqnarray}}
\def\eea{\end{eqnarray}}

\def\NO{\nonumber}

\def\pl#1#2#3{Phys.~Lett.~{\bf B {#1}} ({#2}) #3}
\def\np#1#2#3{Nucl.~Phys.~{\bf B {#1}} ({#2}) #3}
\def\prl#1#2#3{Phys.~Rev.~Lett.~{\bf #1} ({#2}) #3}
\def\pr#1#2#3{Phys.~Rev.~{\bf D {#1}} ({#2}) #3}

\def\ap#1#2#3{Ann.~of Phys.~{\bf {#1}} ({#2}) #3}

\def\ptp#1#2#3{Progr.~Theor.~Phys.~{\bf {#1}} ({#2}) #3}

\renewcommand{\(}{\left(}
\renewcommand{\)}{\right)}

\begin{document}
\date{\mbox{ }}
\title{
{\normalsize DESY 04-121\hfill\mbox{}\\
July 2004\hfill\mbox{}}\\
\vspace{2cm} 
\textbf{The Flavour Puzzle from \\an Orbifold GUT Perspective}\\
[8mm]}
\author{Wilfried Buchm\"uller\\
\\
{\it
Deutsches Elektronen-Synchrotron DESY, Hamburg, Germany}
}
\maketitle

\thispagestyle{empty}

\begin{abstract}
\noindent
Neutrino masses and mixings are very different from quark masses and mixings.
This puzzle is a crucial hint in the search for the mechanism which determines
fermion masses in grand unified theories. We study the flavour problem in 
an $SO(10)$ GUT model in six dimensions compactified on an orbifold.
Three sequential families are localized at three branes where $SO(10)$ is 
broken 
to its three GUT subgroups. Their mixing with bulk fields leads to large 
neutrino mixings as well as small mixings among left-handed quarks. The small 
hierarchy of neutrino masses is due to the mismatch between up-quark and 
down-quark mass hierarchies.
\end{abstract}

\vspace*{2cm}
\noindent
\begin{center}
{\it Talk given at the Fujihara Seminar \\
Neutrino Mass and Seesaw Mechanism, 
KEK, Japan, February 2004 }
\end{center}

\newpage

\section{Gauge unification in six dimensions}

The symmetries and the particle content of the standard model (SM)
point towards grand unified theories (GUTs) as the next step in
the unification of all forces. Left- and right-handed quarks and leptons
can be grouped in three $SU(5)$ multiplets \cite{gg74},  
\bea
{\bf 10} = (q_L,u_R^c,e_R^c)\;, \quad 
{\bf 5^*} = (d_R^c,l_L)\;, \quad {\bf 1}= \n_R^c \;,
\eea
or, alternatively, in two $SU(4)\times SU(2)\times SU(2)$ multiplets 
\cite{ps74}, 
\bea
({\bf 4}, {\bf 2}, {\bf 1}) = (q_L,l_L)\;, \quad  
({\bf 4^*}, {\bf 1}, {\bf 2}) = (u_R^c,d_R^c,\n_R^c,e_R^c)\;.
\eea
All quarks and leptons of one generation are unified in a single multiplet in 
the GUT group $SO(10)$ \cite{gfm75},
\bea
{\bf 16} = {\bf 10} + {\bf 5^*} + {\bf 1} = 
({\bf 4}, {\bf 2}, {\bf 1}) + ({\bf 4^*}, {\bf 1}, {\bf 2}) \;.
\eea
The group $SO(10)$ contains two different $SU(5)\times U(1)$ subgroups,
corresponding to ordinary and `flipped' $SU(5)$ \cite{bar82}, where 
right-handed 
up- and down-quarks are interchanged, yielding another viable 
GUT group. Together with the seesaw mechanism \cite{yan79}, whose twenty-fifth 
anniversery is celebrated at this symposium, grand unified theories provide
an attractive extension of the standard model, which can also account for the
observed smallness of neutrino masses. 

In ordinary four-dimensional (4D) grand unified models, the breaking of the 
GUT symmetry groups to the standard model group
$G_{SM}=SU(3)\times SU(2)\times U(1)$ requires a complicated Higgs sector,
and considerable effort is needed to achieve the wanted gauge symmetry
breaking together with a description of fermion masses and mixings 
that is consistent with experimental data.

Higher-dimensional theories offer new possibilities for gauge symmetry
breaking in connection with the compactification to four dimensions. A
simple and elegant scheme, leading to chiral fermions in four dimensions,
is the compactification on orbifolds, first considered in string theories 
\cite{dhx85,inx87}, and recently revived in the context of effective field
theories in higher dimensions \cite{kaw00}. Orbifold compactifications
lead generically to `split multiplets', i.e. incomplete representations
of the underlying GUT symmetry, which provides a natural mechanism to
split the light weak doublet from the heavy colour triplet Higgs fields.
In the following we shall consider a supersymmetric $SO(10)$ model in
6D \footnote{For $SO(10)$ models in 5D see Refs.~\cite{dm01}.} and discuss
its flavour structure \cite{abc03}.   

Consider now the gauge theory with symmetry group $SO(10)$ in 6D with $N=2$ 
supersymmetry. The gauge
fields $V_M(x,y,z)$, with $M=\m,5,6$, $x^5=y$, $x^6=z$, and the 
gauginos $\l_1$, $\l_2$ are conveniently grouped into vector and chiral
multiplets of the unbroken $N=1$ supersymmetry in 4D,
\beq
V = (V_\m,\l_1)\;, \quad \S = (V_{5,6},\l_2)\;.
\eeq 
Here $V$ and $\S$ are matrices in the adjoint representation of $SO(10)$.

\begin{figure}
\centering 
\includegraphics[scale=0.35]{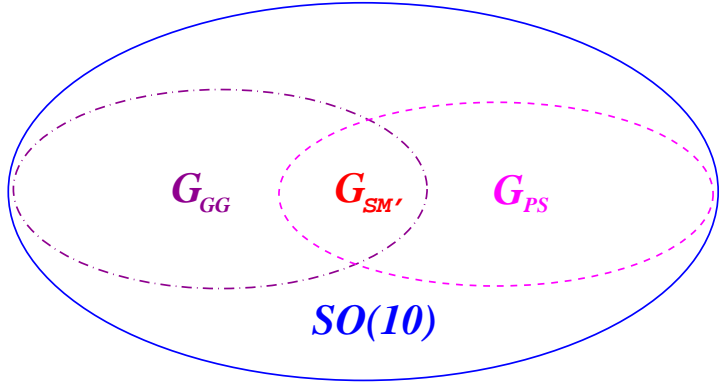}
\caption{The extended standard model gauge group 
$G_{SM'}=SU(3)\times SU(2)\times U(1)^2$ as intersection of the two
symmetric subgroups of $SO(10)$, 
$G_{GG} = SU(5)\times U(1)$ and 
$G_{PS} = SU(4)\times SU(2)\times SU(2)$.}
\label{fig:is}
\end{figure}

Symmetry breaking is achieved by compactification on the
orbifold $T^2/(Z_2^I\times Z_2^{PS}\times Z_2^{GG})$. The discrete symmetries
$Z_2$ break the extended $N=2$ supersymmetry to $N=1$, and the $SO(10)$ gauge 
group to the GUT subgroups (cf.~Fig.~(\ref{fig:is})) 
\bea
G_{PS}=SU(4)\times SU(2)\times SU(2)\;, \quad 
G_{GG}=SU(5)\times U(1)_X \;.
\eea
These breakings are localized at different points in the extra dimensions,
$O=(0,0)$, $O_{PS}=(\p R_5/2,0)$ and $O_{GG}=(0,\p R_6/2)$, where
\bea\label{fixp}
P_IV(x,-y,-z)P_I^{-1} &=& \h_I V(x,y,z)\;,\\
P_{PS}V(x,-y+{\p R_5/2},-z)P_{PS}^{-1} &=& 
\h_{PS} V(x,y+{\p R_5/2},z)\;,\\
P_{GG}V(x,-y,-z+{\p R_6/2})P_{GG}^{-1} &=& 
\h_{GG} V(x,y,z+{\p R_6/2})\;.
\eea 
Here $P_I=I$, the matrices $P_{PS}$ and $P_{GG}$ are given in 
Ref.~\cite{abc01}, and the parities are chosen as $\h_I=\h_{PS}=\h_{GG}=+1$. 
The extended supersymmetry is broken by choosing in the corresponding 
equations for $\S$  all parities $\h_i=-1$. 

\begin{figure}[t]
  \begin{center}
    \centering
    \includegraphics[scale=0.6]{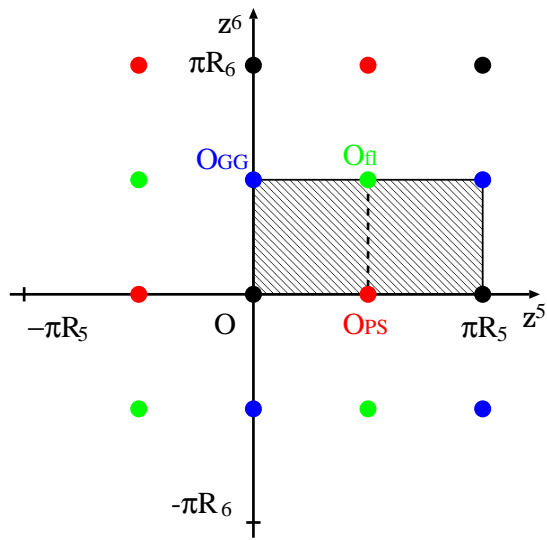}
    \caption{$T^2/(Z_2\times Z_2^{PS} \times Z_2^{GG})$ orbifold
      on the $y$-$z$ plane.  Four different types of fixed points,
      $O$, $O_{PS}$, $O_{GG}$, and $O_{FL}$,
      are denoted by black, red, blue, and green dots, respectively.
      The physical region is, for example, taken as the two-sided rectangle
      formed by folding the shaded region along the dotted line and then
      gluing together the touching edges.}
    \label{fig:cs}
  \end{center}
\end{figure}

There is a fourth fixpoint at  $O_{fl}=(\p R_5/2,\p R_6/2)$, 
which is obtained by combining the three discrete symmetries
$Z_2$, $Z_2^{PS}$ and $Z_2^{GG}$ defined above,
\beq
P_{fl}V(x,-y +{\p R_5/2},-z+{\p R_6/2})P_{fl}^{-1} = 
+ V(x,y,+{\p R_5/2},z+{\p R_6/2})\;.
\eeq 
The unbroken subgroup at the fixpoint $O_{fl}$ is flipped $SU(5)$, i.e.
$G_{fl}=SU(5)'\times U(1)'$. The physical region is obtained by
folding the shaded regions in Fig.~\ref{fig:cs} along the dotted line and 
gluing the edges. The result is a `pillow' with the four fixpoints as corners. 
The unbroken gauge group of the effective 4D theory
is given by the intersection of the $SO(10)$ subgroups at the fixpoints. In
this way one obtains the standard model group with an additional $U(1)$ factor,
\bea
G_{SM'}= SU(3)\times SU(2)\times U(1)_Y \times U(1)_X \;. 
\eea
The difference of
baryon and lepton number is the linear combination
$B-L = \sqrt{{16\over 15}}Y-\sqrt{{8\over 5}}X$. The zero modes of 
the vector multiplet $V$ form the gauge fields of G$_{SM'}$.

The vector multiplet $V$ is a {\bf 45}-plet of $SO(10)$, which has an
irreducible anomaly in 6D. It is related to the irreducible
anomalies of hypermultiplets in the fundamental and the spinor 
representations,
\beq
a({\bf 45}) = - 2 a({\bf 10})\;, \quad 
a({\bf 16}) = a({\bf 16^*}) = - a({\bf 10})\;.
\eeq
Hence, the cancellation of the irreducible anomalies requires two
{\bf 10} hypermultiplets, $H_1 = (H_1,H_1^c)$ and $H_2=(H_2,H_2^c)$,
where the brackets denote the $N=1$ field content. 

The parities for $H_1$ and $H_2$ may be chosen in such a way that one
obtains the Higgs doublets of the MSSM as zero modes,
\bea
H_1^c = H_d\;, \quad H_2 = H_u\;.
\eea
The D-term of the scalar potential has the flat direction
$\langle H_1^c \rangle = \langle H_2 \rangle = v$, which breaks
$SU(2)\times U(1)$ to $U(1)_{em}$. The scale $v$ of electroweak symmetry
breaking can be related to supersymmetry breaking, like in models
of gaugino mediation.

The breaking of $B-L$ can be achieved by adding two {\bf 16}
hypermultiplets, $\Phi^c$ and $\Phi$, with zero modes $N^c$, $N$. The scalar
potential has the D-flat direction
\bea
\langle N^c \rangle = \langle N \rangle = v_N \;, 
\eea
where $v_N \gg v$ can be fixed by a brane superpotential. Anomaly 
cancellation now requires two additional {\bf 10} hypermultiplets $H_3$ and 
$H_4$. The corresponding colour triplet zero modes ($D,G^c$) and ($D^c,G$) 
aquire masses ${\cal O}(v_N)$ from the same brane superpotential \cite{abc02}.

\section{Flavour mixing and seesaw mechanism}

So far we have constructed a gauge theory with symmetry group $SO(10)$ 
and $N=2$ supersymmetry in 6D, locally broken at four fixpoints, such
that the effective low energy theory in 4D has $N=1$ supersymmetry
and the extended SM gauge symmetry $G_{SM'}$. In addition to the vector
multiplet we have two {\bf 16} and four {\bf 10} hypermultiplets. The
corresponding zero modes allow further symmetry breaking of 
$SU(2)\times U(1)$ and $U(1)_{B-L}$ by the ordinary Higgs mechanism. 

How can matter be introduced?  As our guiding principles we shall
use anomaly cancellation and the embedding of quantum numbers in the
adjoint representation of $E_8$\footnote{Such an approach has previously been
pursued in the context of supersymmetric $\s$-models \cite{blx82}.}. 
This implies that only two more {\bf 16} hypermultiplets are allowed,
far too little to account for three quark-lepton generations.
Hence, quarks and leptons must be brane fields. 
As an example, place $\j_1$ at $O_{GG}$, $\j_2$ at $O_{fl}$ and $\j_3$ at 
$O_{PS}$. Here $\j_1$, $\j_2$ and $\j_3$ correspond, up to mixings, to the 
first, second and third generation, respectively. We shall see that this 
assignment is in fact supported by fermion mass relations. 

The three sequential generations are separated by distances large compared 
to the 6D cutoff scale $M_*$. Hence, they can only have diagonal Yukawa 
couplings with the bulk Higgs fields, since direct mixings are exponentially 
suppressed. On the contrary, brane fields can mix with bulk zero modes without
suppression. The embedding into $E_8$ allows two additional {\bf 10} 
hypermultiplets $H_5$, $H_6$, together with two {\bf 16}'s, $\phi$ and 
$\phi^c$. They lead to a partial fourth family together with an anti-family
of zero modes, with the quantum numbers of lepton doublets and down-quark 
singlets,
\bea
L = \left(\begin{array}{l} \n_4 \\ e_4 \end{array}\right)\;, \quad  
L^c = \left(\begin{array}{l} \n^c_4 \\ e^c_4 \end{array}\right)\;, \quad  
G^c_5 = d^c_4\;, \quad G_6 = d_4\;.
\eea
Mixings take place only among left-handed leptons and right-handed 
down-quarks, which is similar to `lopsided' $SU(5)$ models with $U(1)$ 
family symmetry \cite{sy98}. This leads to a characteristic 
pattern of mass matrices. 

Masses and mixings are determined by brane superpotentials. Allowed terms
are restricted by R-invariance and an additional $U(1)_{\tilde{X}}$ symmetry. 
$H_1$, $H_2$, $\F$ and $\F^c$, which aquire a vacuum expectation value, have 
R-charge zero. All matter fields have R-charge one. The {\bf 16}-plets 
$\j_i$ and $\f$ form a quartet $(\j_\a) = (\j_i,\f)$, $\a= 1\ldots4$. The 
brane superpotential reads, for normalized bulk fields, up to quartic 
interactions (23 terms),
\bea\label{O10}
W &=&  M^l_\a \j_\a \f^c  
+ {1\over 2} h^{(1)}_{\a\b} \j_\a \j_\b H_1 
+ {1\over 2} h^{(2)}_{\a\b} \j_\a \j_\b H_2 \NO\\ 
&& + {1\over 2}{h^N_{\a\b}\over M_*}\j_\a\j_\b\F^c\F^c + \ldots \;.
\eea
Here $M_* > 1/R_{5,6} \sim \L_{\rm GUT}$ is the cutoff of the 6D theory.
On the different branes the Yukawa couplings $h^{(1)}$ and $h^{(2)}$ split 
into $h^d, h^e$ and $h^u, h^D$, respectively. 

The breaking of $B-L$ yields masses ${\cal O}(v_N)$ for colour triplet bulk 
zero modes. After electroweak symmetry breaking, 
$\langle H^c_1 \rangle = v_1$, $\langle H_2 \rangle = v_2$,
all zero modes aquire mass terms, and one obtains
\bea
W &=& d_\a m^d_{\a\b} d^c_\b + e^c_\a m^e_{\a\b} e_\b + n^c_\a m^D_{\a\b} \n_\b \NO\\
&& + u^c_i m^u_{ij} u_j + {1\over 2} n^c_i m^N_{ij} n^c_j\;.
\eea
Here $m^d$, $m^e$ and $m^D$ are $4\times 4$ matrices, for instance,
\bea\label{me}
m^e = \left(\begin{array}{cccc}
h^d_{11}v_1 & 0 & 0 & h_{14}^e v_1 \\
0 & h^e_{22}v_1 & 0 & h_{24}^e v_1 \\
0 & 0 & h^d_{33}v_1 & h_{34}^e v_1 \\
M^l_1 & M^l_2 & M^l_3 & M^l_4 \end{array}\right)\;,
\eea
whereas $m^u$ and $m^N$ are diagonal $3\times 3$ matrices,
\bea\label{muN}
m^u = \left(\begin{array}{ccc}
h^u_{11}v_2 & 0 & 0 \\
0 & h^u_{22}v_2 & 0 \\
0 & 0 & h^u_{33}v_2 \end{array}\right)\;, \quad
m^N = \left(\begin{array}{ccc}
h^N_{11}{v_N^2\over M_*} & 0 & 0 \\
0 & h^N_{22}{v_N^2\over M_*} & 0 \\
0 & 0 & h^N_{33}{v_N^2\over M_*} \end{array}\right)\;.
\eea
Some of the mass matrix elements are equal due to GUT relations on the 
corresponding brane, such as $m^e_{11} = m^d_{11}$, $m^D_{22}=m^u_{22}$,
whereas $m^e_{22} \neq m^d_{22}$ (flipped $SU(5)$ brane).

Assuming universal Yukawa couplings at each fixpoint, one obtains a simple 
pattern of quark and lepton mass matrices, significantly different from
4D $SO(10)$ models \cite{alb03}, which reads after a rescaling with the 
appropriate vauum expectation values ($\tan{\b} = v_2/v_1 \simeq 50$), 
\bea
{1\over \tan{\b}}\ m^u \sim {v_1 M_*\over v_N^2}\ m^N \sim 
\left(\begin{array}{ccc}
\m_1 & 0 & 0 \\
0 & \m_2 & 0 \\
0 & 0 & \m_3 \end{array}\right)\;,
\eea
\bea
{1\over \tan{\b}}\ m^D \sim m^e \sim m^d \sim
\left(\begin{array}{cccc}
\m_1 & 0 & 0 & \tm_1 \\
0 & \m_2 & 0 & \tm_2 \\
0 & 0 & \m_3 & \tm_3 \\
\tM_1 & \tM_2 & \tM_3 & \tM_4 \end{array}\right)\;,
\eea
where $\m_i,\tm_i = {\cal O}(v_1)$ and $\tM_i = {\cal O}(\L_{\rm GUT})$. 
The quark-lepton mass spectrum requires $\m_i,\tm_i$ to be hierarchical, 
which may be related to the different location of the three families in the
extra dimension, although this is not explained by the present model.
The GUT mixings $\tM_i$ are assumed to be non-hierarchical.

The parameters $\m_1$, $\m_2$, $\m_3$ are given by the up-quark masses,
which also determine the heavy neutrino masses,
\bea\label{diag}
\m_1 : \m_2 : \m_3 \sim m_u : m_c : m_t \sim M_1 : M_2 : M_3 \;.
\eea
Consider now the down-quark masses and CKM mixings for large 
$\tan{\b} = v_2/v_1 \simeq 50$, such that $h^d_{33} \simeq h^u_{33}$. Since
the hierarchy of down-quark masses is much smaller the one of up-quarks,
it must be dominated by the off-diagonal elements $\tm_i$,
\bea
\m_1 \ll \tm_1\;, \quad  \m_2 \ll \tm_2\;, \quad \m_3 \sim \tm_3\;.
\eea 
These parameters can be fixed by two masses and the Cabibbo angle,
\bea
m_b \simeq \tm_3 \;, \quad m_s \simeq \tm_2 \;, \quad
V_{us}=\Q_c \sim {\tm_1\over \tm_2}\;.
\eea
One then obtains three predictions, the other two mixing angles,
\bea
V_{cb} \sim {m_s\over m_b} \simeq 2\times 10^{-2} \;,\quad
V_{ub} \sim \Q_c {m_s\over m_b} \simeq 4\times 10^{-3} \;,
\eea
and the down quark mass,
\bea
{m_d\over m_s} \sim \g\ \Q_c  \simeq\ 0.03\;, \quad
\g \equiv {\m_2 \over \tm_2} \sim {m_c m_b\over m_t m_s} \sim 0.1 \;,
\eea
which are consistent with data within factor of two. The charged lepton masses
can also be correctly described. The unsuccessful $SU(5)$ relations
$m_s = m_\mu$, $m_d = m_e$ are avoided since the second family is located
at the flipped $SU(5)$ brane.

The neutrino mass matrix can now be computed based on the seesaw mechanism
\cite{yan79},
\bea\label{seesaw}
\bar{m}_\n = - \bar{m}^{D\top} {1\over m^N} \bar{m}^D\;.
\eea
Here $\bar{m}^{D}$ is the $3\times 3$ matrix which is obtained from $m^D$
after integrating out the heavy fourth generation.

The structure of the charged lepton and the Dirac neutrino mass matrices 
is the same. Both matrices lead to large mixings between the `left-handed' 
states. However, due to seesaw mechanism, there is a mismatch between the
matrices which diagonalize the Majorana neutrino and the charged lepton
mass matrices, and a large MNS mixing matrix remains. In a basis where $m^e$ 
is hierarchical, with small off-diagonal terms, the Majorana neutrino 
mass matrix $\bar{m}_\n$ has the form
\bea
\bar{m}_\n \sim \left(\begin{array}{lll}
\g^2 & \g & \g \\
\g & 1 & 1 \\
\g & 1 & 1 \end{array}\right)m_3\;.
\eea
This matrix is familiar from lopsided $SU(5)$ models with $U(1)$ flavour 
symmetry
\cite{sy98}, and it is know to yield a successful neutrino phenomenology.
Note that the small parameter is now determined by quark masses,
$\g \sim m_c m_b/(m_t m_s)$. One characteristic predictions is a large 1-3 
mixing angle, $\Q_{13} \sim \g \sim 0.1$. The coefficients ${\cal O}(1)$ are 
consistent with `sequential heavy neutrino dominance' ($N_3$) \cite{kin00},
yielding large 2-3 mixing, $\sin{2\Q_{23}}\sim 1$.

With $m_3 \simeq \sqrt{\D m_{atm}^2} \sim m_t^2/ M_3$  the heavy Majorana 
masses are $M_3 \sim 10^{15}$~GeV, $M_2 \sim 3\times 10^{12}$~GeV and
$M_1 \sim  10^{10}$~GeV. Decays of $N_1$ may be the origin of the baryon 
asymmetry of the universe \cite{fy86}. In addition to $M_1$, the relevant
quantities are the CP-asymmetry $\ve_1$ and the effective neutrino mass 
$\widetilde{m}_1 = (m^{D\dagger}m^D)_{11}/M_1$. One easily obtains 
$\ve_1 \sim 0.1\; M_1/M_3 \sim 10^{-6}$ and $\widetilde{m}_1 \sim 0.2\; m_3$. 
These are the typical parameters of thermal leptogenesis \cite{bp96}.

One of the most puzzling questions in flavour physics is: Why are masses 
and mixings of quarks and neutrinos so different, and how does this happen in 
grand unified theories where quarks and leptons belong to the same multiplet?
In the context of the discussed 6D $SO(10)$ model these questions have a 
simple answer:
\begin{itemize}

\item The MNS mixings are large because neutrinos are mixtures of brane and 
bulk states, which are unrelated to quark and charged lepton masses and
therefore not suppressed by small mass ratios.

\item The CKM mixings are small because left-handed down-quarks are pure brane
states. The large mixings of right-handed down-quarks, together with the 
down-quark mass hierarchy, leads to small mixings for left-handed down-quarks.

\item Neutrinos have a small mass hierarchy because of the seesaw mechanism 
and the mass relations $m^d \sim m^D$, $m^u \sim m^N$; the `squared' down-quark
hierarchy is almost canceled by the larger up-quark hierarchy,
\bea
{m_1 \over m_3} \sim \left({m_d\over m_b}\right)^2 {m_t\over m_u} \sim 0.1 \;.
\eea

\end{itemize}

The basic mechanism determining the flavour structure is the mixing of
three complete quark-lepton families, localized at three different branes,
with an imcomplete family (split multiplets) originating from the bulk.
This leads to welcome deviations from the familiar GUT mass relations.
The mass hierarchies have a `geometric origin' and are not explained by 
abelian or non-abelian flavour symmetries.

\section{Proton decay}

The 6D $SO(10)$ GUT model makes characteristic predictions for proton decay.
Like in 5D orbifold GUTs, dimension-5 operators are absent \cite{kaw00}.
The dimension-6 operators have an interesting flavour structure due to the
quark-lepton `geography' in the extra dimensions \cite{nom02,hm02}.

In our 6D model the first generation is localized on the $SU(5)$ brane. This
leads to the dimension-6 operators \cite{bcx04}
\bea
{\cal L}_{\rm eff} &=& - {g^2 \over (M_{\cal X}^{\rm eff})^2}\; 
\epsilon_{\a\b\g}
\left(e_1^c u^c_{1 \a} \overline{q}_{1 \b}\overline{q}_{1 \g}
  - d_{1 \a}^c u^c_{1 \b} \overline{q}_{1 \g} \overline{l}_1 \right) \; ,
\eea
where
\begin{eqnarray}
  \frac{1}{(M_{\cal X}^{\rm eff})^2}
  &=& 2 \sum_{n,m=0}^\infty \frac{1}{ M_{\cal X}^2 (n,m)} \NO\\
  &=& 2 \sum_{n,m=0}^{\infty} \frac{R_5^2}{(2 n +1)^2 +
    \frac{R_5^2}{R_6^2} (2m)^2}\;
  \label{sumKK}
\end{eqnarray}
accounts for the sum over all Kaluza-Klein modes; $R_5$ and $R_6$ are the
radii of the two compact dimensions. The sum is logarithmically 
divergent and depends on the cutoff $M_* \sim 10^{17}$~GeV.  
In the symmetric case, $R_5 = R_6 = 1/M_c$, one finds 
\begin{equation}\label{can}
  \frac{1}{(M_{\cal X}^{\rm eff})^2} \simeq \frac{\pi}{
    4\,M_c^2}\(\ln \(\frac{M_*}{M_c}\) + 2.3 \) \; .
\end{equation}

\begin{table}[t]
  \centering
  \begin{tabular}{c|cc|c}
    decay channel & \multicolumn{3}{c}{Branching Ratios [\%]} \\
    \cline{2-4}
    & \multicolumn{2}{c|}{6D {\sf SO(10)}} &
    ${\sf SU(5)}\times {\sf U(1)}_F$ \\
    & case I & case II &  models A \& B\\
    \hline
    $e^+\pi^0$ & 75 & 71 & 54 \\
    $\mu^+\pi^0$ & 4 & 5 & \textless\,1 \\
    $\bar\nu\pi^+$ & 19 & 23 & 27 \\
    $e^+ K^0$ & 1 & 1 & \textless\,1 \\
    $\mu^+ K^0$ & \textless\,1 & \textless\,1 & 18 \\
    $\bar\nu K^+$ & \textless\,1 & \textless\,1 & \textless\,1 \\
    $e^+\eta$ & \textless\,1 & \textless\,1 & \textless\,1 \\
    $\mu^+\eta$ & \textless\,1 & \textless\,1 & \textless\,1 \\
  \end{tabular}
   \caption{Proton decay branching ratios in a 6D $SO(10)$ model compared
    with 4D $SU(5)\times U(1)_F$ models.  
   \label{tb:result}} 
\end{table}

The proton decay branching ratios depend on the overlap of the $SU(5)$ brane 
states 
with the mass eigenstates. Given the mass matrices of Section~2, the 
diagonalization can be explicitly carried out, which leads to the branching
ratios listed in the table \cite{bcx04}. Case I and case II refer to two
different sets of ${\cal O}(1)$ coefficients; the $SO(10)$ results are 
compared with the contribution from the dimension-6 operator in 4D
$SU(5)$ models with $U(1)$ flavour symmetry.

The most striking difference is the decay channel $p\to \mu^+ K^0$,
which is suppressed by about two orders of magnitude in the 6D model
with respect to the 4D models. In both cases the dominant decay mode is 
$p\to e^+ \pi^0$. This is different from 5D $SU(5)$ orbifold GUTs where
the dominant decay modes are $p\to K^+ \nu$ and
$p\to K^0 \m^+$ \cite{nom02,hm02}. 

Finally, a limit on the compactification scale can be derived from the
decay width of the dominant channel $p\to e^+ \pi^0$. One finds 
($M_* = 10^{17}$~GeV),
\begin{equation}
  \Gamma(p\to e^+ \pi^0) 
  \simeq \(\frac{9\times 10^{15}\ {\rm GeV}}{M_c}\)^4
  \(5.3 \times 10^{33}\ {\rm yrs}\)^{-1}\;.
\end{equation}
Hence, the current SuperKamiokande limit $\tau\geq 5.3\times 10^{33}$ 
\cite{SK} yields the lower bound on the compactification scale 
$M_c \geq M_c^{\rm min} \simeq 9\times 10^{15}\,{\rm GeV}$, which is very 
close to the 4D GUT scale 
$\Lambda_{\rm GUT} \simeq 2\times 10^{16}\ {\rm GeV}$ suggested
by the unification of gauge couplings. The choice 
$M_c = \Lambda_{\rm GUT}$ yields the proton lifetime $\t(p\rightarrow
e^+\p^0) \simeq 1\times 10^{35}$~yrs which, remarkably, lies within
the reach of the next generation of large volume detectors!

\section{Towards E$_8$ in higher dimensions}

It is a remarkable group theoretical fact that the adjoint representation
$\bf 45$ of $SO(10)$, together with the spinors $\bf 16$, 
$\bf \overline{16}$ and a $U(1)$ factor form the adjoint representation
of $E_6$. Coset spaces of $E_7$ and $E_8$ with subgroups containing $SO(10)$ 
have similar properties.
This raises the question whether the discussed 6D $SO(10)$ model can be
understood as part of a higher-dimensional theory with gauge group $E_8$,
as it emerges in string theory \cite{gsw87}.

Consider the chain of $E_8$ subgroups
\begin{equation}
E_8 \supset SO(16) \supset SO(10)\times SO(6) \supset \ldots ,
\end{equation}
and the corresponding decomposition of the adjoint representation,
\bea
{\bf 248} &=& {\bf 120} + {\bf 128} \NO\\
&=& {\bf (45,1)} + {\bf (1,15)} + {\bf (10,6)} + 
    {\bf (16,4)} + {\bf (\overline{16},\overline{4})}\; .
\eea
As discussed in Section~2, the 6D $SO(10)$ model has $N=2$ supersymmetry. The 
$\bf 45$ vector multiplet of $SO(10)$, and the six $\bf 10$ and four $\bf 16$
hypermultiplets represent the maximal number of fields from the adjoint
of $E_8$ for which the irreducible 6D anomalies cancel. For this
particular set of fields, also the reducible 6D anomalies cancel \cite{abc03}.

Orbifold GUT models of the type discussed in this talk can occur as
intermediate step in an orbifold compactification of the heterotic
string \cite{krz04,fnx04}. The model of Ref.~\cite{krz04}
\begin{figure}[h]
\centering 
\includegraphics[scale=0.9]{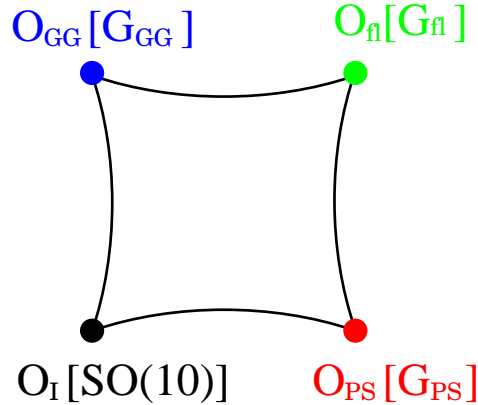}
\caption{The three $SO(10)$ subgroups at the corresponding fixpoints of the
orbifold $T^2/(Z_2^I\times Z_2^{PS}\times Z_2^{GG})$. At each of these
fixpoints one quark-lepton family is localized. \label{fig:orb}}
\end{figure}
indeed contains a 6D $SO(10)$ GUT with the set of bulk fields described
above. At the orbifold fixpoints states from the twisted sector of the
string are localized, as illustrated in Fig.~\ref{fig:orb} for the 
considered orbifold GUT.

Orbifold GUTs have many phenomenologically attractive features. In particular
split multiplets successfully explain the doublet-doublet splitting of 
Higgs fields. Their mixing with brane fields also leads to flavour physics 
with unexpected features and the needed deviations from the simplest GUT 
mass relations.

A successful ultraviolet completion in compactifications of the heterotic 
string will allow to address many questions which go beyond orbifold GUTs.
These include quantum numbers and localization of brane fields, which are
related to twisted sectors of the string, and the unification with gravity. \\

\noindent
I would like to thank T.~Asaka, L.~Covi, D.~Emmanuel-Costa and S.~Wiesenfeldt
for a fruitful collaboration, A.~Hebecker for comments on the manuscript
and the organizers of the seesaw symposium at KEK, especially K.~Nakamura, 
for arranging an enjoyable and stimulating meeting.

\end{document}